\documentclass[conference]{IEEEtran}
\usepackage[left=1.62cm,right=1.62cm,top=1.9cm]{geometry}
\IEEEoverridecommandlockouts

\usepackage{amsmath,amssymb,amsfonts}
\usepackage{algorithmic}
\usepackage{graphicx}
\usepackage{textcomp}
\usepackage{bmpsize}
\usepackage{xcolor}
\usepackage{lipsum}
\usepackage{multirow}
\usepackage{subcaption}
\usepackage{acronym}
\usepackage{enumitem}
\usepackage{comment, float, balance, tabularx, pifont}
\usepackage[ruled,vlined,linesnumbered]{algorithm2e}
\usepackage{hyperref} 
\usepackage[normalem]{ulem}
\usepackage{tikz}
\usepackage{cite, hyperref}
\usepackage{listings}
\usepackage{caption} 
\usetikzlibrary{chains,shapes.multipart}
\usetikzlibrary{shapes,calc}
\usetikzlibrary{automata,positioning}

\SetCommentSty{mycommfont}
\useunder{\uline}{\ul}{}
\def\BibTeX{{\rm B\kern-.05em{\sc i\kern-.025em b}\kern-.08em
    T\kern-.1667em\lower.7ex\hbox{E}\kern-.125emX}}

\definecolor{backcolour}{rgb}{0.95,0.95,0.92}  
\definecolor{codeblue}{rgb}{0.13,0.13,0.68}    
\definecolor{codegreen}{rgb}{0,0.5,0}          
\definecolor{codepurple}{rgb}{0.58,0,0.82}     
\definecolor{codegray}{rgb}{0.5,0.5,0.5}       
\definecolor{codenumbers}{rgb}{0.4,0.4,0.4}    

\lstdefinestyle{pythoncode}{
    commentstyle=\color{codegray},
    keywordstyle=\color{codeblue}\bfseries,
    stringstyle=\color{codepurple},
    identifierstyle=\color{codegreen},
    numberstyle=\tiny\color{codenumbers},
    basicstyle=\ttfamily\small,
    breakatwhitespace=false,         
    breaklines=true,                 
    captionpos=b,                    
    keepspaces=true,                 
    numbers=left,                    
    numbersep=10pt,                  
    showspaces=false,                
    showstringspaces=false,
    showtabs=false,                  
    tabsize=4,
    frame=single,                    
    xleftmargin=15pt,
    xrightmargin=15pt,
    aboveskip=10pt,
    belowskip=10pt,
}

\definecolor{eclipseStrings}{RGB}{42,0.0,255}
\definecolor{eclipseKeywords}{RGB}{127,0,85}
\colorlet{numb}{magenta!60!black}

\lstdefinelanguage{json}{
    basicstyle=\fontsize{10}{10}\normalfont\ttfamily,
    breaklines=true,
    commentstyle=\color{eclipseStrings}, 
    stringstyle=\color{eclipseKeywords}, 
    numbers=left,
    numberstyle=\scriptsize,
    stepnumber=1,
    numbersep=8pt,
    captionpos=b,
    showstringspaces=false,
    xleftmargin=0.5cm,
    xrightmargin=0.5cm,
    frame=single,
    columns=fullflexible,
    aboveskip=10pt,
    belowskip=10pt,
    string=[s]{"}{"},
    comment=[l]{:\ "},
    morecomment=[l]{:"},   
    literate=
        *{0}{{{\color{numb}0}}}{1}
         {1}{{{\color{numb}1}}}{1}
         {2}{{{\color{numb}2}}}{1}
         {3}{{{\color{numb}3}}}{1}
         {4}{{{\color{numb}4}}}{1}
         {5}{{{\color{numb}5}}}{1}
         {6}{{{\color{numb}6}}}{1}
         {7}{{{\color{numb}7}}}{1}
         {8}{{{\color{numb}8}}}{1}
         {9}{{{\color{numb}9}}}{1}
}

\definecolor{backcolour}{rgb}{0.95,0.95,0.92}  
\definecolor{textcolor}{rgb}{0.1,0.1,0.1}      
\definecolor{keywordcolor}{rgb}{0,0,0.6}       
\definecolor{stringcolor}{rgb}{0.58,0,0.82}    

\lstdefinestyle{plaintext}{
    captionpos=b,
    basicstyle=\ttfamily\small\color{textcolor},  
    breaklines=true,                 
    keepspaces=true,                 
    showstringspaces=false,
    frame=single,                    
    xleftmargin=0pt,                 
    xrightmargin=0pt,                
    aboveskip=10pt,
    belowskip=10pt,
    columns=fullflexible,            
    linewidth=\textwidth,            
    keywordstyle=\color{keywordcolor}\bfseries,  
    stringstyle=\color{stringcolor},  
}

\graphicspath{{Figures/}}
\acrodef{3GPP}{3rd Generation Partnership Project}
\acrodef{AI}{Artificial Intelligence}
\acrodef{API}{Application Programming Interface}
\acrodef{AST}{Abstract Syntax Tree}
\acrodef{AWS}{Amazon Web Services}
\acrodef{BERT}{Bidirectional Encoder Representations From Transformers}
\acrodef{CSV}{Comma Separated Values}
\acrodef{DFSDT}{Depth-First Search-Based Decision Tree}
\acrodef{DL}{Deep Learning}
\acrodef{EBS}{Elastic Block Store}
\acrodef{EC2}{Elastic Computing Cloud}
\acrodef{FAISS}{Facebook AI Similarity Search}
\acrodef{FFN}{Feedforward Neural Network}
\acrodef{FFT}{Full Fine-Tuning}
\acrodef{FLO}{Floating-Point Operation}
\acrodef{GB}{Gigabyte}
\acrodef{GenAI}{Generative Artificial Intelligence}
\acrodef{GPT}{Generative Pre-Trained Transformer}
\acrodef{GPU}{Graphics Processing Unit}
\acrodef{HTTP}{Hypertext Transfer Protocol}
\acrodef{JSON}{JavaScript Object Notation}
\acrodef{LLM}{Large Language Model}
\acrodef{LoRA}{Low-Rank Adaptation}
\acrodef{MHM}{Masked Head Model}
\acrodef{ML}{Machine Learning}
\acrodef{NEF}{Network Exposure Function}
\acrodef{NF}{Network Function}
\acrodef{NLP}{Natural Language Processing}
\acrodef{OAS}{OpenAPI Specification}
\acrodef{PaLM}{Pathways Language Model}
\acrodef{PEFT}{Parameter-Efficient Fine-Tuning}
\acrodef{QLoRA}{Q-Low-Rank Adaptation}
\acrodef{RAG}{Retrieval Augmented Generation}
\acrodef{REST}{Representational State Transfer}
\acrodef{SBA}{Service-Based Architecture}
\acrodef{SBI}{Service-Based Interface}
\acrodef{SFT}{Supervised Fine-Tuning Trainer}
\acrodef{SOAP}{Simple Object Access Protocol}
\acrodef{TB}{Terabyte}
\acrodef{URL}{Universal Resource Locator}
\acrodef{XML}{Extensible Markup Language}
\acrodef{YAML}{Yet Another Markup Language}

\begin{document}

\title{NEFMind: Parameter-Efficient Fine-Tuning of Open-Source LLMs for Telecom APIs Automation}

\author{
	\IEEEauthorblockN{
        Zainab Khan\IEEEauthorrefmark{1}\IEEEauthorrefmark{2}, Ahmed Hussain\IEEEauthorrefmark{2}, Mukesh Thakur\IEEEauthorrefmark{3}, Arto Hellas\IEEEauthorrefmark{1}, and Panos Papadimitratos\IEEEauthorrefmark{2}}
	\IEEEauthorblockA{\IEEEauthorrefmark{1}Alto University, School of Science, Espoo, Finland\\
	\IEEEauthorrefmark{2}Networked Systems Security (NSS) Group --
    KTH Royal Institute of Technology, Stockholm, Sweden \\
    \IEEEauthorrefmark{3}Ericsson, Finland
    \\
    Corresponding author: ahmed.hussain@ieee.org
    }
 }

\maketitle

\begin{abstract}
The use of Service-Based Architecture in modern telecommunications has exponentially increased Network Functions (NFs) and Application Programming Interfaces (APIs), creating substantial operational complexities in service discovery and management. We introduce \textit{NEFMind}, a framework leveraging parameter-efficient fine-tuning of open-source Large Language Models (LLMs) to address these challenges. It integrates three core components: synthetic dataset generation from Network Exposure Function (NEF) API specifications, model optimization through Quantized-Low-Rank Adaptation, and performance evaluation via GPT-4 Ref Score and BertScore metrics. Targeting 5G Service-Based Architecture APIs, our approach achieves 85\% reduction in communication overhead compared to manual discovery methods. Experimental validation using the open-source Phi-2 model demonstrates exceptional API call identification performance at 98-100\% accuracy. The fine-tuned Phi-2 model delivers performance comparable to significantly larger models like GPT-4 while maintaining computational efficiency for telecommunications infrastructure deployment. These findings validate domain-specific, parameter-efficient LLM strategies for managing complex API ecosystems in next-generation telecommunications networks.
\end{abstract}

\begin{IEEEkeywords}
Generative AI, Large Language Models, Telecom, Network Exposure Function, 5G, Parameter-Efficient Fine-Tuning
\end{IEEEkeywords}

\section{Introduction}
\label{sec:intro}
The evolution of 5G networks has fundamentally transformed how network services are designed and deployed, particularly through the adoption of the \ac{SBA}~\cite{5g-service-based-core-network-design}. Standardized by the \ac{3GPP}, \ac{SBA} has fundamentally altered how network functionalities are exposed and managed through \acp{NF}. An important element is the \ac{NEF}, which facilitates the exposure of network capabilities through a REST \ac{API}. However, the increase of these \acp{API}, coupled with their increasing complexity, presents significant operational challenges for system administrators and developers.

The challenge manifests itself in two ways. First, the exponential growth in the number of \acp{API} accompanying new \acp{NF} and underlying functionalities has created a vast and complex ecosystem that is increasingly difficult to navigate. Second, the technical intricacies involved in correctly identifying and implementing appropriate \ac{API} calls require substantial expertise and time investment, often leading to inefficiencies and potential errors in system operations. Traditionally, system administrators analyze user queries, navigate through extensive \ac{API} documentation, and formulate appropriate \ac{API} calls. This process is not only time-consuming but also prone to human error, potentially impacting service reliability and system performance. We address these challenges through the use of \acp{LLM}. Specifically, we explore the potential of fine-tuning open-source \ac{LLM}, namely phi-2, to automate the process of interpreting user queries and generating appropriate \ac{NEF} \ac{API} calls.

\textbf{Contribution.} In this paper, we present three key contributions: First, we develop \textit{NEFMind}, a framework encompassing model fine-tuning utilizing \ac{QLoRA} and generating synthetic datasets derived from API specifications using Generative AI. Second, we evaluate model performance through multiple metrics, including GPT-4 Ref Score for accuracy assessment and BertScore for response similarity measurement, providing a reliable evaluation for understanding the capabilities and limitations of our framework. Finally, we demonstrate that fine-tuned \acp{LLM} can effectively automate interactions with \acp{API} while maintaining high accuracy and computational efficiency. Our experimental results show significant improvements in API call automation accuracy, with a fine-tuned model achieving 98-100\% accuracy compared to 4-10\% in the baseline (i.e., non-fine-tuned) Phi-2 model.

\textbf{Paper Organization.} The remainder of this paper is organized as follows: Section~\ref{sec:preliminaries} discusses the preliminaries and relevant related work of API automation and \ac{LLM} applications. Section~\ref{sec:methodology} describes our methodology and proposed framework (\textit{NEFMind}), including the technical implementation of our fine-tuning approach. Section~\ref{sec:performance} provides an in-depth performance evaluation, analyzing the quality of response via correctness and accuracy metrics. Section~\ref{sec:comparison-with-existing-research} highlights the key distinctions between this study and prior research efforts. Finally, Section~\ref{sec:conclusion} concludes the work with directions for future research.
\section{Preliminaries and Related Work}
\label{sec:preliminaries}

\subsection{Web Services and REST APIs in 5G Networks}
This \ac{3GPP}-standardized framework facilitates network functionality exposure through \acp{NF}~\cite{network-exposure}, utilizing \ac{REST} principles for distributed communication~\cite{rest}. REST architecture's core principles—uniform interface, client-server decoupling, statelessness, cacheability, layered systems, and code on demand—establish the foundation for standardized HTTP-based communication while ensuring architectural separation~\cite{rest, rest-api-cs-comm}. In the telecommunications context, these principles are implemented within \ac{NEF} \acp{API}, enabling secure network capability exposure~\cite{nef}. This approach is particularly essential within 5G \ac{SBA}~\cite{SBI-perspective}, as illustrated by the CAMARA project~\cite{camara-project} that demonstrates the evolution toward microservice-based \ac{API} architectures in telecommunications.

\subsection{Large Language Models}
\acp{LLM} fundamentally rely on the Transformer architecture~\cite{transformer-model-architecture}, which revolutionized \ac{NLP} through parallel sequence processing and contextual understanding via self-attention mechanisms~\cite{transformer-bert-gpt}. This architecture integrates multi-head attention layers, position-wise feed-forward networks, residual connections, and layer normalization to enable sophisticated language processing capabilities. Our research examines Phi-2~\cite{phi-2}, an open-source model featuring 2.7 billion parameters and a 2048-token context window that demonstrates efficient performance through training on diverse datasets encompassing Python code and educational content.

\textbf{Model Adaptation Techniques.} Two primary adaptation paradigms have emerged for domain-specific optimization: \ac{RAG} and Fine-Tuning. \ac{RAG} enhances model responses through dynamic knowledge integration from external sources, while fine-tuning~\cite{domain-specialized-LLMs-fine-tuning} directly modifies model weights for specialized domain performance. Zhang et al.~\cite{scaling-meets-llm-fine-tuning} demonstrates the critical relationship between data quality, model architecture, and training methodology in determining fine-tuning effectiveness.

\textbf{Parameter-Efficient Fine-Tuning for Automation.} PEFT techniques~\cite{PEFT-Study} have emerged as highly effective approaches for specialized task adaptation without computational overhead. \ac{QLoRA}~\cite{qlora-paper} represents a significant advancement in this domain, enabling efficient model fine-tuning by updating only specific parameter subsets rather than complete model weights. This approach maintains performance comparable to full fine-tuning while dramatically reducing computational requirements, particularly excelling in code understanding and domain-specific tasks.

\textbf{Integration of \acp{LLM} with APIs.} Contemporary frameworks demonstrate diverse methodological approaches to \ac{LLM}-\ac{API} integration with varying performance characteristics. RestGPT~\cite{restgpt} achieves 70-75\% success rates through a three-module framework architecture, while Gorilla~\cite{gorilla} focuses on comprehensive \ac{API} collections via self-instruction fine-tuning and retrieval-aware training methodologies. ToolLLM~\cite{tool-llm} provides enhanced generalization capabilities for comprehensive \ac{API} integration, and ReST Meets ReAct~\cite{rest-meets-react} advances the field through self-improving agent architectures combining retrieval-based approaches with reactive planning.
\section{Methodology and Proposed Framework}
\label{sec:methodology}
To address the challenges of automating NEF API interactions, we define several key requirements derived from the practical constraints of enterprise telecommunications systems and the need for reliable API automation: (i) Utilization of open-source LLMs to promote research reproducibility and extensibility, (ii) Development of comprehensive in-context knowledge about NEF APIs for accurate query processing, and (iii) Creation of optimal training datasets from API specifications.

\begin{figure}[h!]
    \centering
    \includegraphics[width=0.98\linewidth]{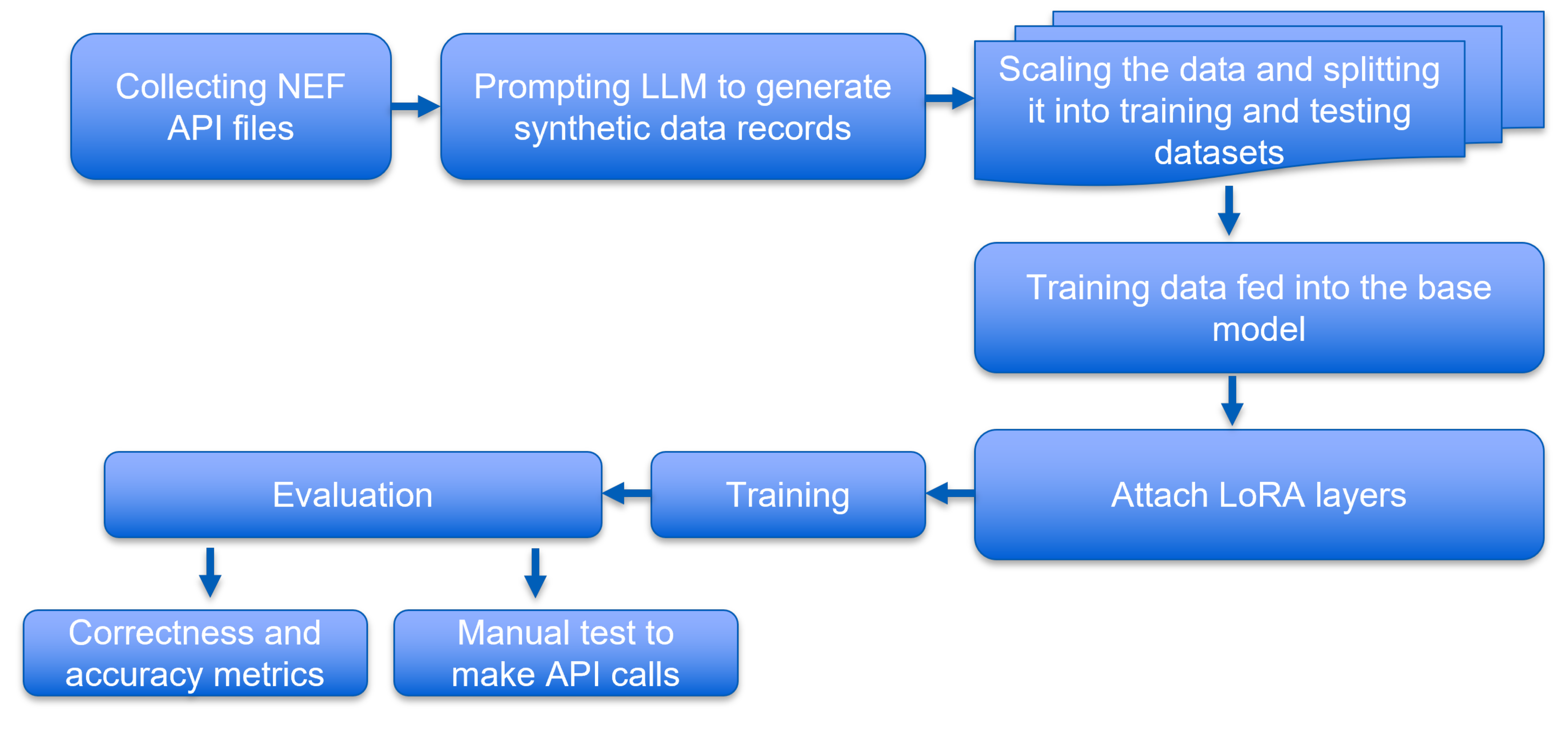}
    \caption{Implementation pipeline consisting of multiple processes
such as synthetic data generation and scaling, fine-tuning the \acp{LLM} and their evaluation.}
    \label{fig:implementation-pipeline}
\end{figure}

\textbf{Experimental Setup.} We utilized Amazon G5 instances to evaluate \textit{NEFMind}. Specifically, our setup employed the g5.4xlarge instance configuration for comprehensive model operations, encompassing data generation, training, and evaluation of the Phi-2 architecture. The g5.4xlarge configuration consists of a single GPU with 24GB of dedicated GPU memory, 16 virtual CPUs, and 64GB of system memory. This instance provides 600GB storage capacity with network and EBS bandwidth capabilities of 25Gbps and 8Gbps, respectively.

\textbf{Design and Framework Implementation.} The proposed methodology addresses the aforementioned requirements through a systematic approach incorporating model fine-tuning of state-of-the-art language model--Phi-2--utilizing a specialized dataset derived from the \ac{NEF} \ac{API} specification documentation~\cite{5GC_APIs}. To establish comprehensive performance benchmarks, we conduct a comparative analysis of the fine-tuned model and a \ac{RAG} architecture implemented with GPT-4. The implementation adopts a modular pipeline architecture, where each component functions as an integral part of a sequential processing chain, with output artifacts serving as input parameters for subsequent processing stages. As illustrated in Fig.~\ref{fig:implementation-pipeline}, building the framework encompasses four primary components: the initial dataset collection and synthetic data generation process, followed by data scaling and preprocessing operations, subsequent model fine-tuning utilizing QLoRA, and concluding with systematic evaluation and performance assessment protocols. NEFMind specifically addresses telecommunications infrastructure constraints through parameter-efficient adaptation, requiring 60\% less computational resources than comparable full fine-tuning approaches while maintaining domain-specific accuracy.

\textbf{Dataset Generation.} The effective integration of telecommunication \ac{API} knowledge into \acp{LLM} necessitates a specialized fine-tuning process utilizing domain-specific datasets. We employ \ac{NEF} \ac{API} specifications~\cite{nef} as the foundational knowledge source for model adaptation. Further, we leverage \ac{NEF} \ac{API} specification files in \ac{YAML} format for synthetic data generation, implementing a systematic approach to dataset development. Below we discuss the methodology encompassing data generation, processing mechanisms, and scaling, establishing a robust foundation for model fine-tuning and subsequent performance evaluation.

\textbf{Generating Synthetic Data.} Initially, we utilized the original dataset~\cite{5GC_APIs} consisting of the YAML specification files of \ac{NEF} \acp{API}. To facilitate model comprehension, we flattened one of the YAML files, as each \ac{API} specification file contains references to other files, with no external references beyond the file itself. Next, we formulated a prompt for the \ac{LLM} to generate synthetic data in \ac{JSON} format, encompassing specific fields including user queries (Request), endpoint specifications (\ac{API} call, Method, Operation), contextual information (Description), and required parameter configurations (Parameters). For this task, we employed GPT-4~\cite{gpt4} as the expert model to ensure the highest quality of data. Although the \ac{API} specification delineates schema definitions for seven \ac{API} endpoints and the prompt specified the return of solely real data, the \ac{LLM} (GPT-4) produced some fabricated and unreal \ac{JSON} objects. 

After data generation, the data validation and refinement are performed. Manual inspection and verification of the GPT-4 generated dataset identifies and eliminates instances of synthetic data artifacts that did not align with actual \ac{NEF} \ac{API} specifications. We isolated seven \ac{JSON} objects that demonstrated complete fidelity to the authentic \ac{API} specifications. This refinement ensures the dataset's validity and establishes a reliable foundation for subsequent model training and evaluation processes. Fig.~\ref{fig:synthetic_json} illustrates the generated \ac{JSON} object after the refinement process.

\begin{figure}[!h]
\begin{minipage}{\linewidth}
\begin{lstlisting}[label={lst:generated_synthetic_json_data}, language=json, basicstyle=\ttfamily\tiny, breaklines=true]
{
    "request": "How can I obtain an access token for future requests?",
    "api_call": "/api/v1/login/access-token",
    "description": "OAuth2 compatible token login, get an access token for future requests",
    "method": "post",
    "operation": "login_access_token_api_v1_login_access_token_post",
    "parameters": {
        "grant_type": "password",
        "username": "string",
        "password": "string",
        "scope": "string",
        "client_id": "string",
        "client_secret": "string"
    }
},
{
    "request": "How can I read active subscriptions?",
    "api_call": "/api/v1/3gpp-as-session-with-qos/v1/{scsAsId}/subscriptions",
    "description": "Get subscription by id",
    "method": "get",
    "operation": "read_active_subscriptions_api_v1_3gpp_as_session_with_qos_v1
    __scsAsId__subscriptions_get",
    "parameters": {
        "scsAsId": "string"
    }
}
\end{lstlisting}
\end{minipage}
\caption{Generated synthetic JSON data for API requests and their required parameters.}
\label{fig:synthetic_json}
\end{figure}

\begin{figure}[!h]
\includegraphics[width=0.8\columnwidth, height=7cm]{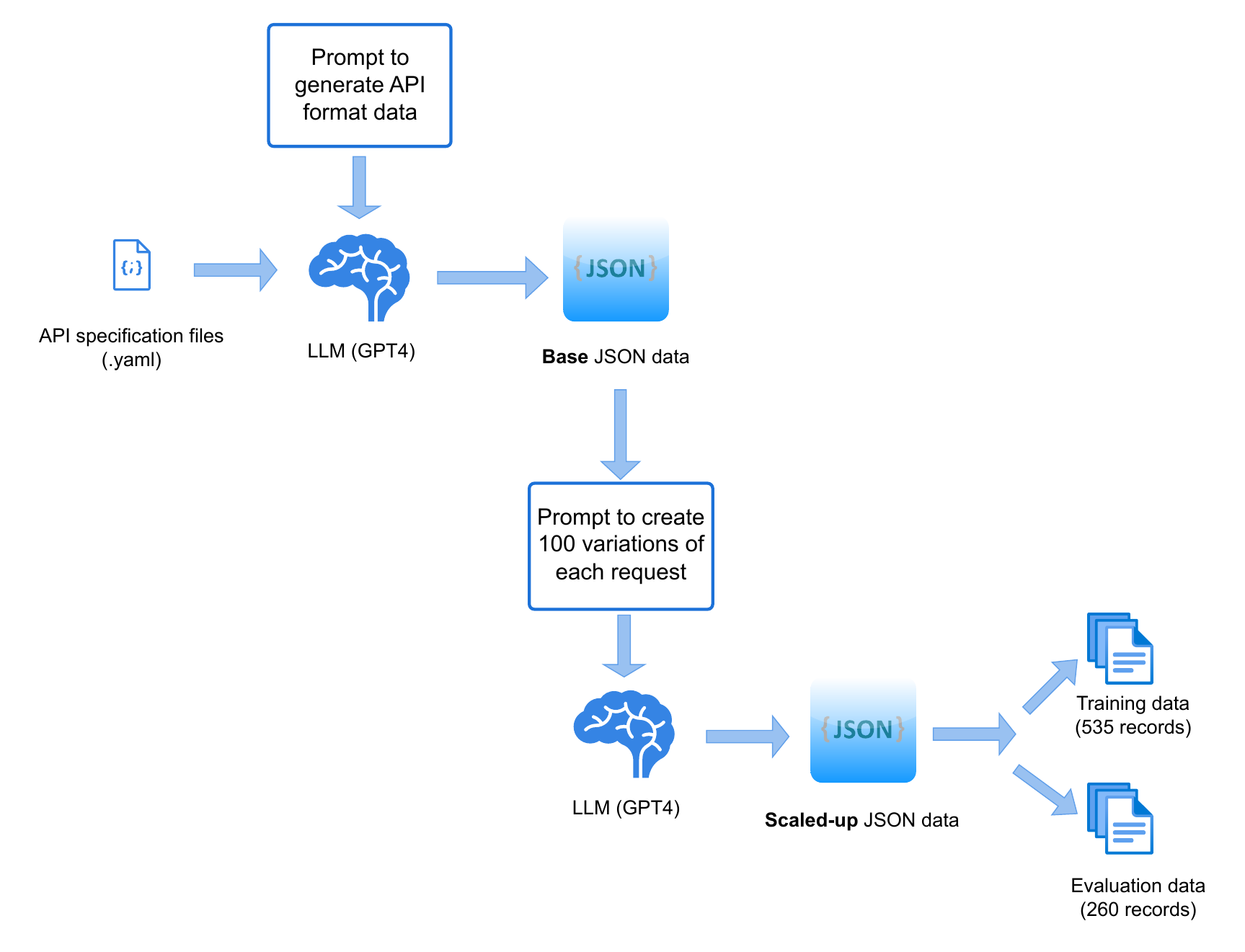}
\centering
\caption{Synthetic data generation pipeline incorporating GPT-4-based question-answer pair generation from \ac{NEF} \ac{API} specifications, iterative data augmentation, and dataset partitioning for model evaluation.}
\label{fig:data_generation_process}
\end{figure}

\textbf{Data Processing and Scaling.} After obtaining seven records of data from the previous step, we recognized the inadequacy of this dataset for effective model training. Consequently, we opted to prompt the GPT-4 model to expand the dataset: We provided the model with the request field of each JSON record and requested it to generate 100 unique variations of each request. This process resulted in the generation of 765 records of data. The prompt for scaling the data is given in Fig.~\ref{fig:data-scaling-prompt}. Subsequently, we partitioned this expanded dataset roughly into a 70/30 ratio for training and evaluation purposes, resulting in approximately 535 records in the training dataset and 230 in the evaluation dataset. To facilitate the model training, we converted the JSON data into Comma Separated Values (CSV) format, as it was deemed suitable for the model trainer’s comprehension. Additionally, for Phi-2, we structured each data record into \textit{Instruct}-\textit{Output} pairs. The Instruct field contained the value from the request parameter, while the Output object encompassed the remaining parameters such as API call, description, method, operation, and parameters. The complete process of data generation is depicted in Fig.~\ref{fig:data_generation_process}.

\begin{figure}[!h]
\centering
\begin{minipage}{0.9\linewidth}
\begin{lstlisting}[style=plaintext]

Understand the given JSON object and generate the request parameter in 100 different ways. All of them must be unique and not redundant.
    {json_object}

You may also use the document provided as context to understand more. Feel free to rephrase the request parameter.
{context}

The format of output must be an array of 100 values in the following format:
[request1, request2, ..., request100]

\end{lstlisting}
\end{minipage}
\caption{Prompt for data scaling.}
\label{fig:data-scaling-prompt}
\end{figure}

\enlargethispage{-\baselineskip}

\textbf{Fine-Tuning Phi-2.} Phi-2 is a Transformer-based architecture comprising 2.7 billion parameters originally trained on a diverse dataset encompassing Python textbook content, coding data, \ac{NLP}-based synthetic texts, and educational website content. We utilize Phi-2 for fine-tuning telecommunication-based \ac{API} datasets, specifically selected for its moderate parameter scale and being open-source, enabling performance analysis in domain-specific applications. The fine-tuning implemented a systematic approach encompassing multiple integrated phases. Initially, the training dataset underwent formatting procedures, incorporating appropriate labeling schemas and implementing the \textit{Instruct/Output} format structure. Following data preparation, the formatted \ac{CSV} dataset was integrated into the processing pipeline. The base Phi-2 model was subsequently initialized without quantization parameters but incorporating FlashAttention-2 optimization, enhancing the model's attention mechanism efficiency for extended sequence processing while optimizing computational resource utilization and memory management. The model initialization phase was followed by tokenization and applying the \ac{QLoRA} configurations. The final implementation phase utilized HuggingFace's \ac{SFT} framework~\cite{SFT-Huggingface}, integrating the prepared model architecture with the specified \ac{QLoRA} configurations (Table~\ref{table:lora_configuration_arguments}) and training parameters (Table~\ref{table:training_arguments}). 

\begin{table}[!h]
\centering
\caption{\ac{QLoRA} Configuration}
\label{table:lora_configuration_arguments}
\resizebox{0.67\columnwidth}{!}{%
\begin{tabular}{|c|c|}
\hline
\textbf{Argument} & \textbf{Value} \\
\hline
\ac{LoRA} alpha & 16 \\
\hline
\ac{LoRA} dropout & 0.1 \\
\hline
\ac{LoRA} rank & 64 \\
\hline
Target modules & q\_proj, k\_proj, v\_proj, dense, fc1, fc2 \\
\hline
Bias & None \\
\hline
Task type & CAUSAL\_LM \\
\hline
\end{tabular}%
}
\end{table}

\begin{table}[!h]
\centering
\caption{Training parameters provided to \ac{SFT} from HuggingFace to fine-tune the model.}
\label{table:training_arguments}
\resizebox{0.67\columnwidth}{!}{%
\begin{tabular}{|c|c|}
\hline
\textbf{Argument} & \textbf{Value} \\
\hline
Epochs & 5 \\
\hline
Batch size & 3 \\
\hline
Gradient accumulation steps & 1 \\
\hline
Optim & paged\_adamw\_32bit \\
\hline
Save steps & 10 \\
\hline
Logging steps & 10 \\
\hline
Learning rate & $2 \times 10^{-4}$ \\
\hline
Weight decay rate & 0.001 \\
\hline
Warmup ratio & 0.03 \\
\hline
BF16 & True \\
\hline
Max grad norm & 0.3 \\
\hline
Max steps & -1 \\
\hline
Group by length & True \\
\hline
Scheduler type & Constant \\
\hline
Reports to & Tensorboard \\
\hline
\end{tabular}%
}
\end{table}

The model achieved a training runtime of approximately 595 seconds, processing 4.495 training samples per second and executing 1.504 training steps per second, with a total of \acp{FLO} of $6.08 \times 10^{15}$. The final training loss of 0.1921 indicates an optimal balance between computational efficiency and model performance, suggesting Phi-2's particular suitability for scenarios prioritizing rapid fine-tuning capabilities despite potentially higher loss metrics compared to alternative architectures.

\section{Performance Evaluation}
\label{sec:performance}
We evaluate the performance of the baseline (i.e., Phi-2) and fine-tuned open-source \ac{LLM}, utilizing two performance metrics: accuracy and response similarity. The evaluation comprises 230 records extracted from a comprehensive dataset of 765 entries~\cite{5GC_APIs}, incorporating three fundamental components: \ac{NEF} \ac{API} endpoint identification queries, GPT-4 reference responses as evaluation benchmarks, and comparative model outputs from baseline and fine-tuned variants. We employ 25 iterative evaluations, enabling granular analysis of the accuracy and similarity of the generated responses. Each evaluation iteration generates structured \ac{JSON} artifacts encompassing prediction datasets with performance metrics, BertScore-based semantic similarity quantification, and GPT-4 Ref Score comparative analysis of accuracy~\cite{karapantelakis2024using} for reference-based benchmarking.

\begin{figure*}[t]
    \centering
    \begin{subfigure}[b]{0.23\textwidth}
        \includegraphics[width=0.98\textwidth]{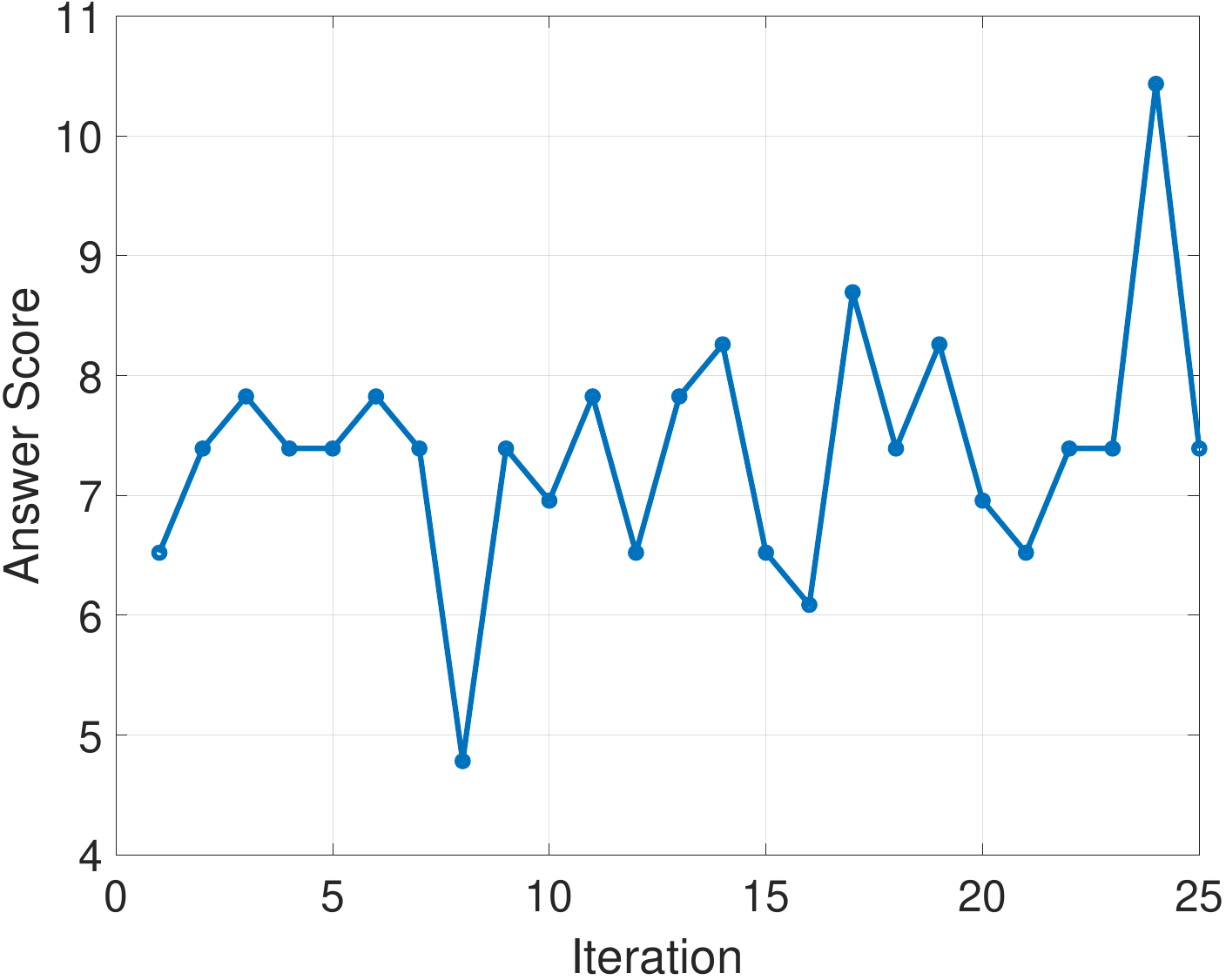}
        \caption{}
        \label{fig:base-phi2-gpt4ref}
    \end{subfigure}
    ~
    \begin{subfigure}[b]{0.23\textwidth}
        \includegraphics[width=0.98\textwidth]{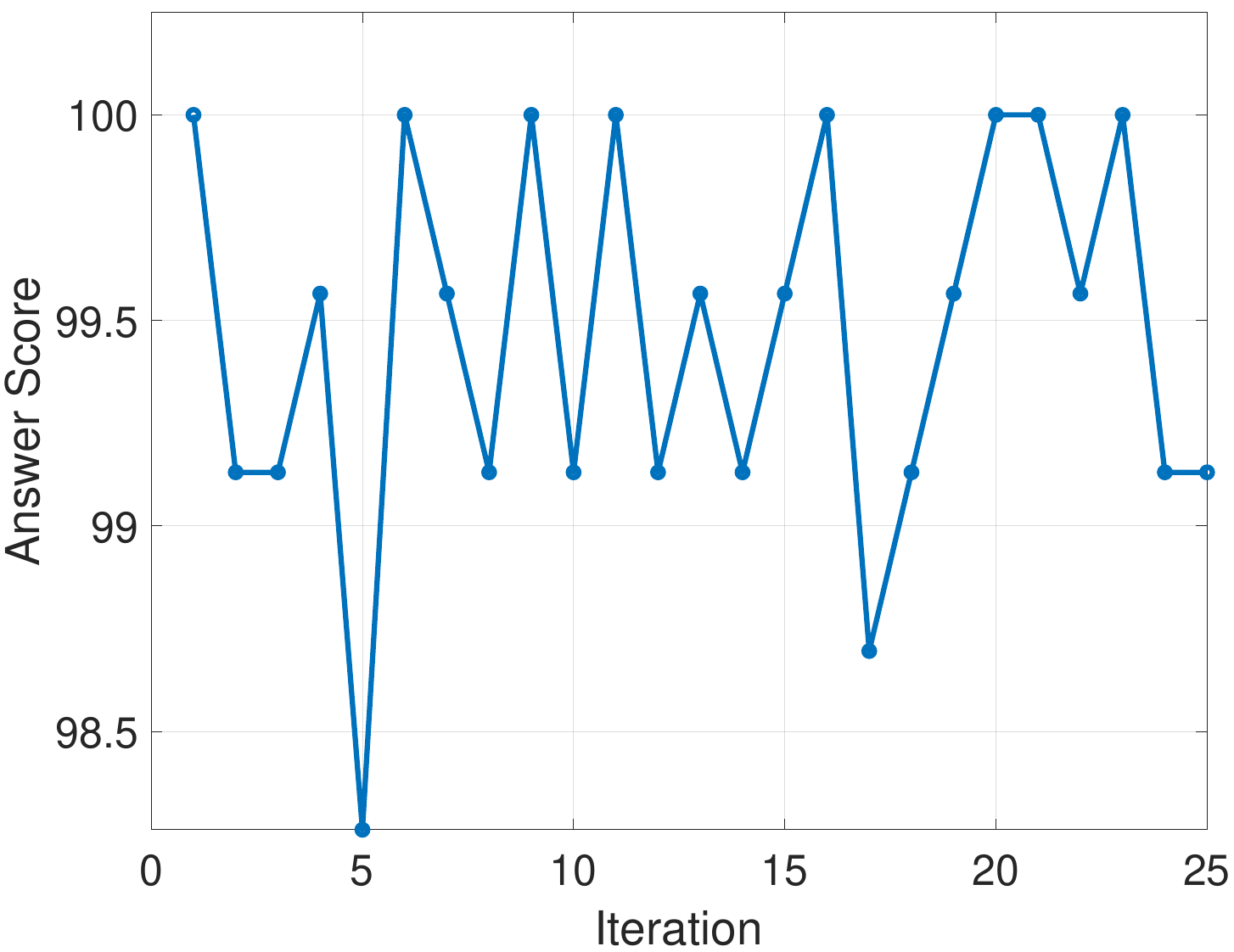}
        \caption{}
        \label{fig:fine-tuned-phi2-gpt4ref}
    \end{subfigure}
    ~
    \begin{subfigure}[b]{0.23\textwidth}
        \includegraphics[width=0.98\textwidth]{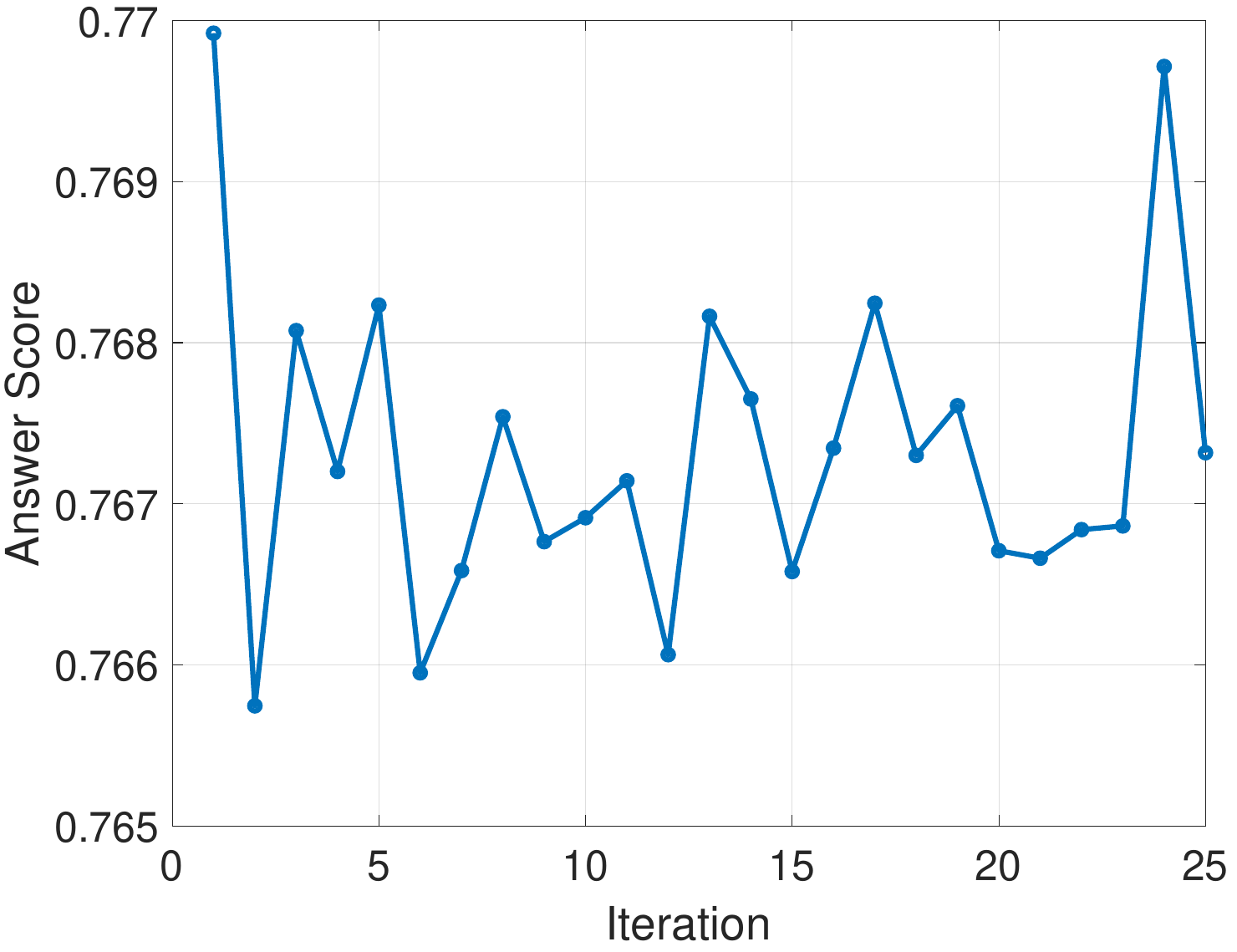}
        \caption{}
        \label{fig:base-phi2-bert-score}
    \end{subfigure}
    ~
    \begin{subfigure}[b]{0.23\textwidth}
        \includegraphics[width=0.98\textwidth]{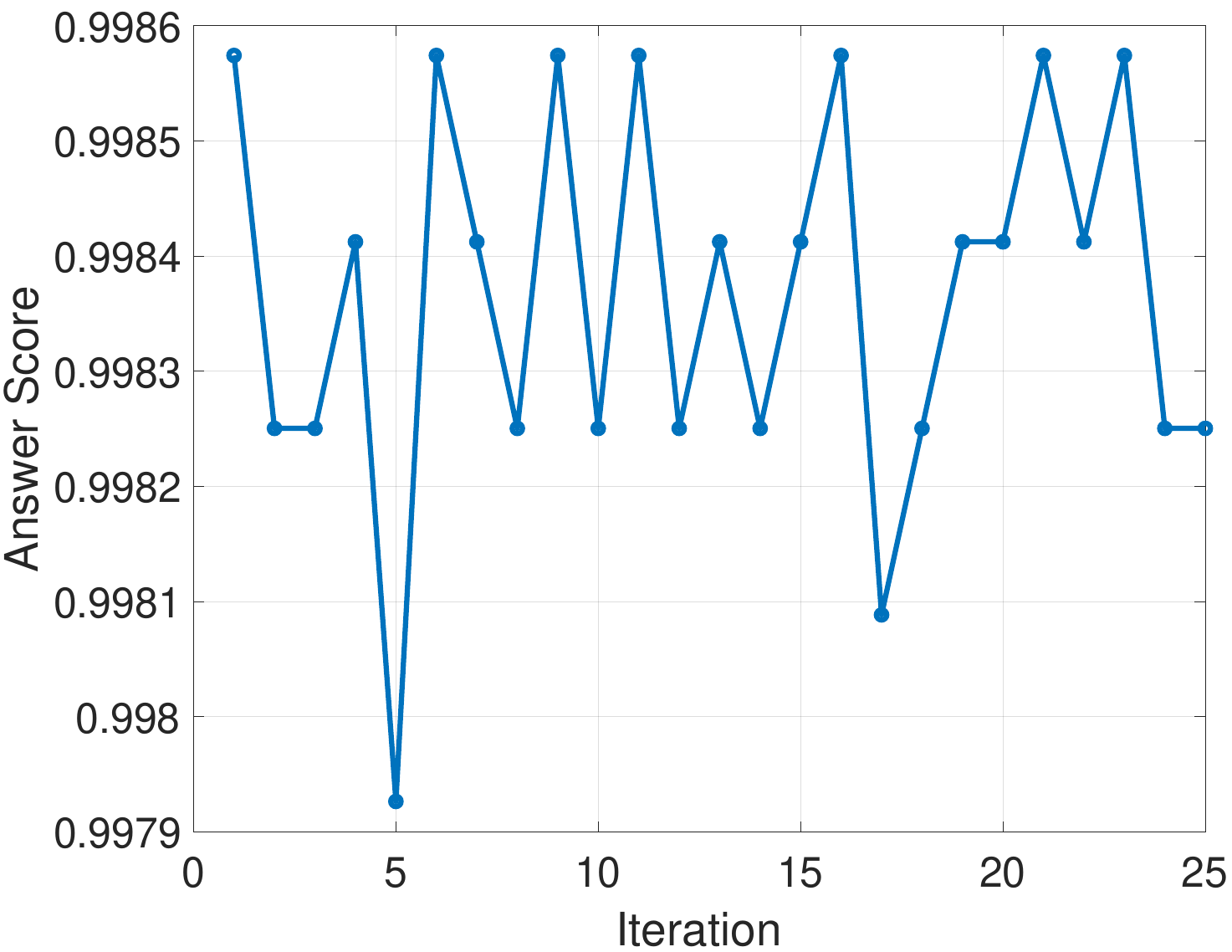}
        \caption{}
        \label{fig:fine-tuned-phi2-bert-score}
    \end{subfigure}
    \caption{Evaluation metrics across 25 iterations comparing base Phi-2 and fine-tuned Phi2-NEF models. (a) GPT-4 Ref Scores for base Phi-2, (b) GPT-4 Ref Scores for Phi2-NEF, (c) BERTScore for base Phi-2, and (d) BERTScore for Phi2-NEF.}
    \label{fig:all_figures}
\end{figure*}

\textbf{Baseline Model Evaluation Framework.} The baseline Phi-2 evaluation implements a \ac{RAG}-based methodology to address inherent domain-specific knowledge limitations. Our processing pipeline systematically integrates \ac{API} specifications through semantic segmentation utilizing LangChain's recursive text splitter~\cite{recursive-text-splitter}, followed by HuggingFace embedding generation and \ac{FAISS}-based similarity retrieval~\cite{faiss-langchain}. The architecture integrates LangChain's Q\&A framework~\cite{question-answer-chain} with task-specific prompt structures, generating evaluation outputs in structured \ac{JSON} format. Analysis reveals significant formatting inconsistencies and semantic inadequacies in baseline model responses, demonstrating the critical necessity for domain-specific adaptation.

\textbf{Fine-Tuned Model Assessment Protocol.} The fine-tuned Phi-2 architecture processes \ac{NEF} \ac{API} queries independently without supplementary contextual information. Systematic evaluation using 230 records validates model performance through HuggingFace pipeline integration with task-specific tokenization. Response formatting protocols convert string-format outputs into structured \ac{JSON} objects, systematically documenting responses under designated keys alongside corresponding inputs and ground truth references.

\textbf{Performance Metrics and Analysis.} Our dual-metric assessment protocol employs GPT-4 Ref Score~\cite{karapantelakis2024using} for accuracy quantification and BertScore for semantic similarity analysis. Each metric executes across 25 iterations, generating comprehensive performance characterization as presented in Table~\ref{stats-of-evaluation-results}.

\textbf{GPT-4 Reference Score Analysis.} Utilizing \ac{GPT}-4~\cite{gpt4, karapantelakis2024using} as the expert validation model, accuracy assessment reveals stark performance disparities. Baseline Phi-2 demonstrates accuracy ranges of 4.78-10.43 on a 0-100 scale, while fine-tuned variants achieve 98.69-100.00 ranges. This substantial improvement validates the effectiveness of domain-specific training methodologies, as illustrated in Figures~\ref{fig:base-phi2-gpt4ref} and~\ref{fig:fine-tuned-phi2-gpt4ref}.

\textbf{BertScore Semantic Similarity Assessment.} BertScore evaluation~\cite{bert-score-paper} employs BERT-based contextual embeddings with HuggingFace frameworks, transforming natural language into vector representations for pairwise cosine-similarity analysis. Baseline Phi-2 achieves similarity scores within 0.765-0.769 ranges on a 0-1 scale, while fine-tuned variants demonstrate 0.997-0.998 ranges. These results, visualized in Figures~\ref{fig:base-phi2-bert-score} and~\ref{fig:fine-tuned-phi2-bert-score}, validate our curated prompting strategy's effectiveness in generating well-structured, consistent \ac{NEF} \ac{API} interaction outputs.

\begin{table}[t]
\caption{Performance metrics summary across twenty-five evaluation iterations for baseline and fine-tuned Phi-2 models.}
\label{stats-of-evaluation-results}
\begin{tabular}{l|c|c|c|c|}
\cline{2-5}
& \multicolumn{2}{c|}{\textbf{GPT-4 Ref Score}} & \multicolumn{2}{c|}{\textbf{BertScore}} \\ \cline{2-5} 
& \textbf{Phi-2} & \textbf{Phi2-NEF} & \textbf{Phi-2} & \textbf{Phi2-NEF} \\ \hline
\multicolumn{1}{|l|}{\textbf{Maximum}} & 10.4348 & 100 & 0.7699 & 0.9986 \\ \hline
\multicolumn{1}{|l|}{\textbf{Minimum}} & 4.7826 & 98.2609 & 0.7657 & 0.9979 \\ \hline
\end{tabular}
\end{table}

\textbf{Empirical \ac{API} Integration Validation.} Despite superior response generation performance, direct \ac{API} integration testing reveals implementation challenges. Validation employs LangChain's OpenAPI agent framework~\cite{create-openapi-agent} with standardized testing credentials, implementing systematic query interpretation and execution planning. The agent architecture, fundamentally based on \ac{OAS}, demonstrates the fine-tuned model's limitations in producing deterministic, semantically consistent outputs within integrated frameworks. These observations highlight potential agent integration incompatibilities and architectural constraints requiring systematic investigation in future research iterations.

Our comprehensive evaluation demonstrates that baseline Phi-2 exhibits significant performance limitations due to insufficient domain-specific knowledge, while fine-tuned architectures achieve superior performance through specialized \ac{NEF} \ac{API} training. However, the limited training corpus scale introduces potential overfitting risks, necessitating expanded datasets in future development cycles to maintain model generalization capabilities.
\section{Comparison with Existing Solutions}
\label{sec:comparison-with-existing-research}
Several notable research efforts have explored the integration of \acp{LLM} with \acp{API}, each contributing distinct methodological approaches and domain applications that provide valuable context for understanding our work's positioning within the field.

RestGPT~\cite{restgpt} represents a foundational approach to \ac{LLM}-\ac{API} integration through its prompt-based framework architecture. This system employs three interconnected modules---a planner, API selector, and executor---designed to interpret complex tasks, decompose them into actionable steps, and execute appropriate RESTful \ac{API} calls across general-purpose applications. While RestGPT demonstrates broad applicability using OpenAI's text-davinci-003 model with sophisticated prompt engineering techniques, our research is different in both domain specificity and methodological approach. We focus exclusively on telecommunications \acp{API}, enabling deeper domain expertise and more targeted optimization. 

Furthermore, our implementation leverages fine-tuning techniques on open-source models rather than relying solely on prompt engineering, ensuring greater accessibility for researchers and independence from proprietary systems. This distinction yields substantial performance improvements: while RestGPT achieved 70-75\% success rates for correct \ac{API} calls, our approach demonstrates 98-100\% prediction accuracy within the telecommunications domain.

Gorilla~\cite{gorilla} explores the integration of \acp{LLM} with RESTful \acp{API}. However, significant technical and evaluative differences distinguish their and our approach. Gorilla employs self-instruction fine-tuning combined with retrieval-aware training techniques, whereas our methodology centers on \ac{PEFT} strategies. Both utilize open-source model foundations; Gorilla refining LLaMA-7B-based architectures while we similarly leverage accessible model resources. The evaluation frameworks differ distinctly: Gorilla directly benchmarks its fine-tuned model against GPT-4, demonstrating superior \ac{API} functionality performance and reduced hallucination errors. 

Conversely, our study employs GPT-4 exclusively as a teacher model, focusing evaluation efforts on comparing fine-tuned models with their baseline counterparts rather than direct comparison with proprietary systems.

ToolLLM~\cite{tool-llm} aligns closely with our objectives through its comprehensive framework for bridging \acp{LLM} with external knowledge via RESTful \ac{API} interactions. Their methodology encompasses complete data construction, model training, and evaluation pipelines. The primary distinction lies in dataset scope and domain strategy: ToolLLM aggregates diverse datasets from RapidAPI Hub spanning 49 categories, pursuing broad multi-domain applicability. 

In contrast, our approach deliberately constrains focus to the telecommunications industry, enabling specialized optimization and targeted performance enhancements within this critical infrastructure domain.

\section{Conclusion}
\label{sec:conclusion}
In this paper, we presented \textit{NEFMind}, a framework for automating \ac{NEF} \ac{API} interactions through domain-specific fine-tuning of open-source \acp{LLM}. Our analysis of Phi-2 architecture, when applying Parameter-Efficient Fine-Tuning, demonstrates significant performance enhancements in \ac{API} call automation. The fine-tuned architecture achieved 98-100\% accuracy (compared to 4-10\% baseline) and improved BertScore metrics from 0.76 to 0.99 while maintaining computational efficiency through \ac{QLoRA} implementation. Notably, the performance of the Phi-2 architecture (2.7B parameters) suggests viable deployment in telecommunications infrastructure. While these findings establish the efficacy of fine-tuning for \ac{API} call generation and demonstrate the viability of parameter-efficient models for domain-specific tasks, challenges persist in security implementation and cross-domain generalization. Future research should address these limitations while enhancing architectural frameworks and fine-tuning methodologies.

\balance
\bibliographystyle{IEEEtran}
\bibliography{main}
\end{document}